\def\lncs{0}
\def\marginnotes{1}
\ifnum\lncs=0
\documentclass[11pt]{article}
\usepackage{fullpage}
\else
\documentclass[runningheads]{llncs}
\fi
\usepackage{amsmath, amssymb, amsfonts, mathrsfs,verbatim,color}
\ifnum\lncs=0
\usepackage{amsthm}
\newtheorem{lemma}{Lemma}
\newtheorem{theorem}{Theorem}
\newtheorem{claim}{Claim}

\theoremstyle{definition}
\newtheorem{definition}{Definition}

\fi

\def\inr{\leftarrow}

\newcommand{\Gen}{\mathsf{Gen}}
\newcommand{\Rep}{\mathsf{Rep}}
\newcommand{\rep}{\mathsf{Rep}}

\newcommand{\M}{\mathcal{M}}
\newcommand{\from}{\leftarrow}
\newcommand{\tm}{\tilde{m}}
\newcommand{\set}[1]{\left\{{#1}\right\}}
\def\ff{\mathsf{SS}}
\def\win{\mathsf{Succ}}
\def\tr{\mathsf{tr}}
\def\sketch{{\sketch}}
\def\entss{\entss}
\def\ssrec{\mathsf{SRec}}

\newcommand{\definitionclose}{\hspace*{\fill}$\diamondsuit$}

\def\l{\ell}
\newcommand{\suchthat}{\,|\,}

\def\bool{\ensuremath{\{0, 1\}}}

\newcommand{\hinf}{{\mathbf{H}_{\infty}}}
\newcommand{\thinf}{{\widetilde{\mathbf{H}}_{\infty}}}
\newcommand{\Ext}{\mathsf{Ext}}
\newcommand{\paren}[1]{\left( #1 \right)}

\newcommand{\sd}[1]{\mathbf{SD}\paren{{#1}}}
\DeclareMathOperator*{\expe}{\mathbf E}
\newcommand{\bydef}{\stackrel{\mbox{\tiny{def}}}{=}}

\newcommand{\dis}[2]{{\mathsf{dis}(#1,#2)}}
\newcommand{\disfn}{\mathsf{dis}}
\newcommand{\logeps}{\log\tfrac{1}{\varepsilon}}
\newcommand{\logdel}{\log\tfrac{1}{\delta}}

\newcommand{\mypar}[1]{{\vspace{.2cm}\noindent \sc #1}}

\def\bool{\{0,1\}}
\newcommand{\bit}[1]{{\{0,1\}}^{#1}}
\def\key{R}  % short-term session key
\def\Rep{\mathsf{Rep}}  %%% use Rep for extractors, no \Rec anywhere!

\def\zo{\{0,1\}}
\def\A{{\cal A}}

\def\M{{\cal M}}

\def\H{{\cal H}}

\def\sS{{\mathsf{S}}}
\def\bad{{\mathsf{Bad}}}
\newcommand{\tuple}[1]{({#1})}
\newcommand{\ignore}[1]{}
\def\to{\rightarrow}

\def\adv{{\A}}

\def\eps{\varepsilon}

\def\seed1{{d_1}}
\def\output1{{l_1}}

\def\F{{\mathbb F}}

\ifnum\marginnotes=0
\newcommand{\lnote}[1]{}
\newcommand{\bnote}[1]{}
\else
\newcommand{\lnote}[1]{\textcolor{red}{\textbf{!!!}}\marginpar{\tiny \sf #1 --L.R.}}
\newcommand{\bnote}[1]{\textcolor{red}{\textbf{!!!}}\marginpar{\tiny \sf #1 --B.K.}}
\fi

\title{An Improved Robust Fuzzy Extractor}
\ifnum\lncs=1
\author{Bhavana Kanukurthi and Leonid Reyzin}
\institute{Boston University Computer Science\\
111 Cummington St., Boston, MA 02215, USA\\
\texttt{http://cs-people.bu.edu/bhavanak, http://www.cs.bu.edu/\~{ }reyzin}}
\authorrunning{Bhavana Kanukurthi and Leonid Reyzin}
\tocauthor{Bhavana Kanukurthi and Leonid Reyzin (Boston U.)}
\titlerunning{An Improved Robust Fuzzy Extractor}
\else
\author{Bhavana Kanukurthi and Leonid Reyzin\\
Boston University Computer Science\\
\texttt{http://cs-people.bu.edu/bhavanak, http://www.cs.bu.edu/\~{ }reyzin}}
\fi

\begin{document}

\maketitle

\begin{abstract}
  We consider the problem of building robust fuzzy extractors, which
  allow two parties holding similar random variables $W$, $W'$ to
  agree on a secret key $R$ in the presence of an active adversary.
  Robust fuzzy extractors were defined by Dodis et al. in Crypto 2006
  to be noninteractive, i.e., only
  one message $P$, which can be modified by an unbounded adversary,
  can pass from one party to the other.  This allows them to be used
  by a single party at different points in time (e.g., for key
  recovery or biometric authentication), but also presents an
  additional challenge: what if $R$ is used, and thus possibly
  observed by the adversary, before the adversary has a chance
  to modify $P$.   Fuzzy extractors secure against such a strong attack
  are called post-application robust.

  We construct a fuzzy extractor with post-application robustness that
  extracts a shared secret key of up to $(2m-n)/2$ bits (depending
  on error-tolerance and security parameters), where $n$
  is the bit-length and $m$ is the entropy of $W$. The
  previously best known result, also of Dodis et al., extracted up to $(2m-n)/3$ bits (depending
  on the same parameters).
\end{abstract}

\section{Introduction}
Consider the following scenario.  A user Charlie has a secret $w$ that
he wants to use to encrypt and authenticate his hard drive.  However, $w$ is not a
uniformly random key; rather, it is a string with some amount of
entropy from the point of view of any adversary $\adv$.  Naturally,
Charlie uses an extractor~\cite{NZ96}, which is a tool for converting
entropic strings into uniform ones.  An extractor $\Ext$ is an
algorithm that takes the entropic string $w$ and a uniformly random
seed $i$, and computes $R=\Ext(w;i)$ that is (almost) uniformly random
even given $i$.

It may be problematic for Charlie to memorize or store the uniformly
random $R$ (this is in contrast to $w$, which can be, for example, a
long passphrase already known to Charlie, his biometric, or a physical token,
such as a physical one-way function~\cite{PRTG02}).  Rather, in order
to decrypt the hard drive, Charlie can use $i$ again to recompute
$R=\Ext(w;i)$.  The advantage of storing $i$ rather than $R$ is that
$i$ need not be secret, and thus can be written, for example, on an
unencrypted portion of the hard drive.  

Even though the storage of $i$ need not be secret, the authenticity of
$i$ is very important.  If $\adv$ could modify $i$ to $i'$, then
Charlie would extract some related key $R'$, and any guarantee on the
integrity of the hard drive would vanish, because typical encryption
and authentication schemes do not provide any security guarantees
under related-key attacks.  To authenticate $i$, Charlie would need to
use some secret key, but the only secret he has is $w$.

This brings us to the problem of building \emph{robust} extractors:
ones in which the authenticity of the seed can be verified at
reconstruction time.  A robust extractor has two procedures: a
randomized $\Gen(w)$, which generates $(R, P)$ such that $R$ is
uniform even given $P$ (think of $P$ as containing the seed $i$ as
well as some authentication information), and $\rep(w, P')$, which
reproduces $R$ if $P'=P$ and outputs $\bot$ with high probability for
an adversarially produced $P'\neq P$. 

Note that in the above scenario, the adversary $\adv$, before
attempting to produce $P'\neq P$, gets to see the value $P$ and how
the value $R$ is used for encryption and authentication.  Because we
want robust fuzzy extractors to be secure for a wide variety of
applications, we do not wish to restrict how $R$ is used and,
therefore, what information about $R$ is available to $\adv$.  Rather,
we will require that $\adv$ has low probability of getting $\rep(w,
P')$ to not output $\bot$ \emph{even} if $\adv$ is given both $P$ and
$R$.  This strong notion of security is known as
\emph{post-application} robustness.

An additional challenge may be that the value $w$ when $\Gen$ is run
is slightly different from the value $w'$ available when $\rep$ is
run: for example, the user may make a typo in a long passphrase, or a
biometric reading may differ slightly.  Extractors that can tolerate
such differences and still reproduce $R$ exactly are called
\emph{fuzzy}~\cite{DORS08}.  Fuzzy extractors are obtained by adding
error-correcting information to $P$, to enable $\rep$ to compensate
for errors in $w'$.  The specific constructions depend on the kinds of
errors that can occur (e.g., Hamming errors, edit distance errors,
etc.).

Robust (fuzzy) extractors are useful not only in the single-party
setting described above, but also in interactive settings, where two
parties are trying to derive a key from a shared (slightly different
in the fuzzy case) secret $w$ that either is nonuniform or about which
some limited information is known to the adversary $\adv$. One party,
Alice, can run $\Gen$ to obtain $(R, P)$ and send $P$ to the other
party, Bob, who can run $\rep$ to also obtain $R$.  However, if $\adv$
is actively interfering with the channel between Alice and Bob and
modifying $P$, it is important to ensure that Bob detects the
modification rather than derives a different key $R'$.  Moreover,
unless Alice can be sure that Bob truly received $P$ before she starts
using $R$ in a communication, post-application robustness is
needed.

\mypar{Prior Work.}
Fuzzy extractors, defined in~\cite{DORS08}, are essentially the 
noninteractive variant of privacy amplification 
and information reconciliation protocols, %which are 
considered in multiple works, including~\cite{Wyner,BBR88,Mau93,BBCM95}. 
Robust (fuzzy) extractors, defined in~\cite{BDKOS05,DKRS06}, 
are the noninteractive variant of privacy amplification 
(and information reconciliation) secure against active 
adversaries~\cite{Mau97,MauW97,Wolf98,MauW03c,RW03,RW04B}.  

Let the length of $w$ be $n$ and the entropy of $w$ be $m$.
Post-application robust fuzzy extractors cannot extract anything out
of $w$ if $m<n/2$, because an extractor with post-application
robustness implies an information-theoretically secure message
authentication code (MAC) with $w$ as the key%
\footnote{The MAC is
  obtained by extracting $R$, using it as a key to any standard
  information-theoretic MAC (e.g.,~\cite{WC81}), and sending $P$ along with the tag
  to the verifier},
which is impossible if $m<n/2$ (see \cite{DS02} for impossibility of deterministic MACs
	if $m<n/2$
and its extension by \cite{Wic08} to randomized MACs). Without any set-up 
assumptions, the only previously
known post-application robust extractor, due to \cite{DKRS06},
extracts $R$ of length $\frac{2}{3}(m-n/2-\logdel)$ (or even less
if $R$ is required to be very close to uniform), where $\delta$ is
the probability that the adversary violates robustness. Making it
fuzzy further reduces the length of $R$ by an amount related to the
error-tolerance. (With set-up assumptions, one can do much better:
the construction of~\cite{CDFPW08}
extracts almost the entire entropy $m$,
reduced by an amount related to security and, in the fuzzy case, to
error-tolerance.  However, this construction assumes that a nonsecret
uniformly random string is already known to both parties, and that the
distribution on $w$, including adversarial knowledge about $w$, is
independent of this string.)

\mypar{Our Results.}
The robust extractor construction of~\cite{DKRS06} is parameterized by
a value $v$ that can be decreased in order to obtain a longer $R$.  In
fact, as shown in~\cite{DKRS06}, a smaller $v$ can be used for \emph{pre-application} robustness
(a weaker security notion, in which $\adv$ gets $P$ but not $R$).  We
show in Theorem~\ref{theorem-dkrs-lowerbound}
that the post-application-robustness analysis of~\cite{DKRS06} is
essentially tight, and if $v$ is decreased, the construction becomes insecure.

Instead, in Section~\ref{sec-rfe-errorless}, we propose a new construction of an extractor with
post-application robustness that extracts $R$ of length
$m-n/2-\logdel$, improving the previous result by a factor of
$3/2$ (more if $R$ is required to be very close to uniform).
While this is only a constant-factor increase, in scenarios where
secret randomness is scarce it can make a crucial difference.  
Like~\cite{DKRS06}, we make no additional set-up
assumptions. Computationally, our construction is slightly more
efficient than the construction of~\cite{DKRS06}.  Our improved robust
extractor translates into an improved robust fuzzy extractor using the
techniques of~\cite{DKRS06}, with the same factor of $3/2$
improvement. 

In addition, we show (in Section~\ref{section-dkrs-uniformity-dominates-improvement})
a slight improvement for the pre-application robust version of the extractor
of~\cite{DKRS06}, applicable when the extracted string must be particularly close
to uniform.

\section{Preliminaries}
\label{section-prelims}
\mypar{Notation.}
For binary strings $a,b$, $a||b$ denotes their concatenation, $|a|$
denotes the length of $a$. For a binary string $a$, for 
we denote by $[a]^j_i$, the substring $b = a_ia_{i+1}\ldots a_j$.  
If $S$ is a set, $x\leftarrow S$ means that $x$ is chosen uniformly 
from $S$. If $X$ is a probability distribution (or a random variable), then $x\leftarrow X$ 
means that $x$ is chosen according to distribution $X$. 
If $X$ and $Y$ are two random variables, then $X\times Y$ denotes
the product distribution (obtained by sampling $X$ and $Y$ 
\emph{independently}). All logarithms are 
base $2$.

\mypar{Random Variables, Entropy, Extractors.}
Let $U_l$ denote the uniform distribution on $\{0,1\}^l$.
Let $X_1, X_2$ be two probability distributions over some set $S$.
Their {\em statistical distance} is 
\[\sd{X_1, X_2} \bydef \max_{T\subseteq S} 
\{ \Pr[X_1 \in T] - \Pr[X_2 \in T] \}=
\frac{1}{2}\sum_{s\in S} \left|\Pr_{X_1}[s]-\Pr_{X_2}[s]\right |\]
(they are said to be $\eps$-close if $\sd{X_1, X_2}\le \eps$).
We will use the following lemma on statistical distance 
that was proven in~\cite{DKRS08}: 
\begin{lemma}
\label{lemma-sdswitch}
For any joint distribution $(A, B)$ and distributions
$C$ and $D$ over the ranges of $A$ and $B$ respectively,
if $\sd{(A, B), C\times D}\le \alpha$, then 
%\ifnum\lncs=0
%$\sd{(A, B), C\times B}\le 2\alpha$.
%\else
$\mathbf{SD}((A,B), \allowbreak C \times B)  \le 2\alpha$.
%\fi
\end{lemma}

\mypar{Min-entropy.} The \emph{min-entropy} of a random variable $W$ is 
\ifnum\lncs=1 defined as 
\fi 
$\hinf(W) = -\log(\max_{w}\Pr[W=w])$ (all logarithms are base 2,
unless 
specified otherwise). 
Following \cite{DORS08}, for a joint distribution $(W, E)$,
define the (average) conditional min-entropy of $W$ 
given $E$ as \[\thinf(W\mid E) =\allowbreak -\log (\expe_{e\inr E}(2^{-\hinf(W\mid E=e)}))\] 
(here the expectation is taken over $e$ for which $Pr[E=e]$ is nonzero).
A computationally unbounded adversary who receives the value of $E$ cannot find the correct value of $W$
with probability greater than $2^{-\thinf(W\mid E)}$. 
We will use the following lemma from ~\cite{DORS08}: 
\begin{lemma}
\label{lemma-avg-min-entropy}
Let $A,B,C$ be random variables. If $B$ has at most $2^\lambda$ possible values, then 
\ifnum\lncs=0
the following holds: 
$\thinf(A|B,C) \allowbreak \geq \thinf((A,B)|C) -\lambda \geq \thinf(A|C) -\lambda$. 
\else
$\thinf(A|B,C) \geq \thinf((A,B)|C) -\lambda \geq \thinf(A|C) -\lambda$. 
\fi
In particular, $\thinf(A|B) \geq \hinf((A,B)) -\lambda \geq \hinf(A) -\lambda$.
\end{lemma}

Because in this paper the adversary is sometimes assumed to have some external
information $E$ about Alice and Bob's secrets, we need the following variant,
defined in~\cite[Definition 2]{DORS08}, of the definition of strong extractors
of~\cite{NZ96}:

\begin{definition}
Let $\Ext:\zo^n \to \zo^l$ be a polynomial time 
probabilistic function that uses $r$ bits of randomness. 
We say that $\Ext$ is an 
{\sf average-case $(n, m, l, \eps)$-strong extractor}
if for all pairs of random variables 
$(W,E)$ such that $w \in W$ is an $n$-bit string and 
$\thinf(W\mid E) \geq m$, we have 
\ifnum\lncs=0
$\sd{(\Ext(W;X), X,E), (U_l,X,E)}\le \eps$,
\else
$\mathbf{SD}((\Ext(W;X), X,E), \allowbreak (U_l,X,E))\le \eps$,
\fi
where $X$ is the uniform distribution
over $\zo^r.$ 
\end{definition}

Any strong extractor can be made average-case with a slight increase in input
entropy~\cite[Section 2.5]{DORS08}. We should note that some strong extractors, 
such as universal hash functions~\cite{CW79,HILL99} discussed next,
generalize without any loss to average-case.  

\mypar{The Leftover Hash Lemma} 
We first recall the notion of universal hashing~\cite{CW79}:
\begin{definition}
A family of efficient functions $\H= \set{h_{i}:\bool^n \rightarrow
\bool^\l}_{i \in I}$ is \emph{universal} if for all
distinct $x, x'$ we have $\Pr_{i \leftarrow I}[h_i(x) = h_i(x')] \leq
2^{-l}$. 

$\H$ is \emph {pairwise independent} if for all distinct 
$x, x'$ and all $y, y'$ it holds that $\Pr_{i \in I}[h_i(x) = y \wedge
h_i(x') = y'] \leq 2^{-2\l}$.  %Families with $\delta = 2^{-\l}$
%are called \emph{pairwise independent}. 
\definitionclose
\end{definition}

\begin{lemma}[Leftover Hash Lemma, average-case version~\cite{DORS08}]
\label{lemma-leftover-hash}
For $\l, m, \eps>0$, $\H$ is a strong $(m,
\eps)$ average-case extractor (where the index of the hash function is
the seed to the extractor)
if $\H$ is universal and $\l \le m+2-2\logeps$.
\end{lemma}
\noindent This Lemma easily generalizes to the case
when $\H$ is allowed to depend on the extra
information $E$ about the input $X$. In other words,
every function in $\H$ takes an additional input $e$, and the family $\H$ is
universal for every fixed value of $e$.

\mypar{Secure Sketches and  Fuzzy Extractors.}\label{sec:fuzzy-defs}
We start by reviewing the
definitions of secure sketches and fuzzy extractors from
\cite{DORS08}. Let $\M$ be a metric space with distance function
$\disfn$ (we will generally denote by $n$ the length of each element in
$\M$).
Informally, a secure sketch enables recovery
of a string $w\in \M$ from any ``close'' string $w'\in \M$ without leaking too
much information about~$w$.

\begin{definition} \label{def:ss}
An {\sf $(m,\tm,t)$-secure sketch} is a pair of efficient
randomized procedures ($\ff,\ssrec$) s.t.: 
\begin{enumerate}
\item The sketching procedure $\ff$ on input $w\in \M$ returns a bit
string $s\in \bit{*}$.
The recovery procedure  $\ssrec$ takes an element $w'\in \M$ and
$s\in \bit{*}$.

\item \emph{Correctness:} If $\dis{w}{w'}\le t$ then $\ssrec(w',\ff(w))=w$.
\item \emph{Security:} For any
distribution $W$ over $\M$ with min-entropy $m$, the (average)
min-entropy of $W$ conditioned on $s$ does not decrease  very
much. Specifically, if $\hinf(W) \geq m$ then $\thinf(W\mid
\ff(W))\geq \tm$.
\end{enumerate}
The quantity $m-\tm$ is called the {\em entropy
loss} of the secure sketch.
\definitionclose
\end{definition}

In this paper, we will construct a robust fuzzy extractor 
for the binary Hamming metric using secure sketches for the 
same metric. 
We will briefly review the syndrome construction from~\cite[Construction 3]{DORS08}
that we use
(see also references therein for its previous incarnations).
Consider an efficiently decodable $[n,n-k,2t+1]$
linear error-correcting code $C$.  The sketch $s=\ff(w)$ consists of
the $k$-bit syndrome $w$ with respect to $C$.  
We will use the fact
that $s$ is a (deterministic) linear function of $w$ and that the
entropy loss is at most $|s|=k$ bits in the construction of our 
robust fuzzy extractor for the Hamming metric. 

We note that, as was shown in \cite{DKRS06}, the secure sketch 
construction for the set difference metric of $\cite{DORS08}$ can
be used to extend the robust fuzzy extractor construction 
in the Hamming metric to the set difference metric. 

While a secure sketch enables recovery of a string $w$ from a close string $w'$,
a fuzzy extractor extracts a close-to-uniform string $\key$ 
and allows the precise reconstruction of $\key$ 
from any string $w'$ close to $w$. 

\begin{definition} \label{def:fe}
An {\sf $(m,\ell,t,\eps)$-fuzzy extractor} is a pair of efficient
randomized procedures ($\Gen,\Rep$) with the following properties:
\begin{enumerate}
\item The generation procedure $\Gen$, on input $w\in \M$, outputs an
extracted string $\key\in \zo^\ell$ and a helper string $P\in
\bit{*}$. The reproduction procedure $\Rep$ takes an element
$w'\in \M$ and a string $P\in \bit{*}$ as inputs.

\item \emph{Correctness:}
If $\dis{w}{w'}\le t$ and 
$(\key, P)
\from \Gen(w)$, then $\Rep(w',P)=\key$.

\item \emph{Security:}
For any distribution $W$ over $\M$ with min-entropy $m$, the
string $\key$ is close to uniform even conditioned on the value
of~$P$. Formally, if $\hinf(W) \geq m$ and $\tuple{\key,P} \from
\Gen(W)$, then we have
$\sd{(\key,P), U_\ell \times P }\le \eps$.
 \definitionclose
\end{enumerate}
\end{definition}

Note that fuzzy extractors allow the information $P$ to be revealed
to an adversary without compromising the security of the 
extracted random string $\key$. However, they provide no guarantee when 
the adversary is active. 
Robust fuzzy extractors defined (and constructed) in ~\cite{DKRS06} 
formalize the notion of security against active adversaries. We review the definition below. 

If $W, W'$ are two (correlated) random variables over a metric space
$\M$,
we say $\dis{W}{W'}\le t$ if
the distance between $W$ and $W'$ is at most $t$ with probability one.
We call $(W,
W')$ a {\em $(t, m)$-pair} if $\dis{W}{W'} \leq t$ and $\hinf(W)
\geq m$. 

\begin{definition}\label{def:robust}
An $(m, \ell, t, \eps)$-fuzzy extractor has {\sf post-application 
\ifnum\lncs=0(resp., pre-application) 
\else (resp., pre-appli- \allowbreak cation)
\fi
    robustness $\delta$} if for all $(t,m)$-pairs $(W,
W')$ and all adversaries $\adv$, the probability that the following
experiment outputs ``success'' is at most $\delta$: sample $(w, w')$ from $(W,
W')$; let $(\key, P) = \Gen(w)$; let $\tilde P = \adv(\key, P)$ (resp.,
$\tilde P = A(P)$); output ``success'' if $\tilde P \neq P$ and $\rep(w',
\tilde{P}) \neq \perp$.
\definitionclose
\end{definition}

We note that the above definitions can be easily extended to give
\emph{average-case} fuzzy extractors (where the
adversary has some external information $E$ correlated with $W$), and that our
constructions satisfy those stronger definitions, as well.

\section{The New Robust Extractor}
\label{sec-rfe-errorless}
In this section we present our new extractor with post-application
robustness.  We extend it to a robust \emph{fuzzy} extractor in
Section~\ref{sec-rfe-errors}.  Our approach is similar to that 
of~\cite{DKRS06}; a detailed comparison is given in
Section~\ref{sec-comparison-dkrs}.

\mypar{Starting point: key agreement secure against a passive adversary.}
Recall that a strong extractor allows extraction of 
a string that appears uniform to an adversary even given 
the presence of the seed used for extraction. 
Therefore, a natural way of achieving key agreement in the
 errorless case is for Alice to pick a random seed $i$
  for a strong extractor and send it to Bob (in the clear). 
They could then use $R=\Ext(w;i)$ as the shared key.
 As long as the adversary is passive, the shared key looks uniform to her. 
 However, such a protocol can be rendered completely insecure 
 when executed in the presence of an active adversary because 
$\adv$ could adversarially modify $i$ to $i'$ such 
that $R'$ extracted by Bob has no entropy. 
To prevent such malicious modification of $i$ we will require 
Alice to send an authentication of $i$ (along with $i$) to Bob.
 In our construction, we authenticate $i$ using $w$ 
 as the key and then extract from $w$ using $i$ 
as the seed. Details follow. 

\mypar{Construction.}
For the rest of the paper we will let $w \in \zo^n$. We will
assume that $n$ is even (if not, drop one bit of $w$, reducing its entropy by at most 1).
To compute $\Gen(w)$, let 
$a$ be the first half of $w$ and $b$ the second:
$a=[w]_1^{n/2}, b=[w]_{n/2+1}^n$.
 View $a$,$b$ as elements of $\F_{2^{n/2}}$. 
Let $v=n-m+\logdel$, where $\delta$ is the desired
robustness.
Choose a random $i \in \F_{2^{n/2}}$.  
Compute $y = ia+b$. 
Let $\sigma$ consist of the first
$v$ bits of $y$ and the extracted key $R$ consist of the rest of $y$: 
$\sigma = [y]^v_1$, $R = [y]^{n/2}_{v+1}.$ 
Output $P = (i,\sigma).$% (See Figure~\ref{f-robust-key-agreement}). 

\begin{tabbing}
\ifnum\lncs=0
wwwwwwwwwwww\=www\=3. \=wwwwwwwww\=\kill
\else
wwwwww\=wwww\=3. \=wwwwwwwww\=\kill
\fi
\>\+
$\Gen(w)$:\\
1. Let $a=[w]_1^{n/2}, b=[w]_{n/2+1}^n$ \\
2. Select a random $i\leftarrow \F_{2^{n/2}}$\\
3. Set $\sigma = [ia+b]^{v}_{1}$, $R= [ia+b]^{n/2}_{v+1}$ and output $P=(i, \sigma)$\\ 
\ \\
$\Rep(w,P' = (i',\sigma'))$:\\
1. Let $a=[w]_1^{n/2}, b=[w]_{n/2+1}^n$ \\
2. If $\sigma' = [i'a+b]^{v}_{1}$ then compute $R' = [i'a+b]^{n/2}_{v+1}$ else output $\bot$
\end{tabbing}

\begin{theorem}
Let $\M = \zo^n.$ Setting $v = n/2 - \ell$, 
the above construction is an $(m,\ell,0,\eps)-\mathsf{fuzzy\ extractor}$ 
with robustness $\delta$, for any $m,\ell,\eps,\delta$ satisfying 
$\ell \leq m - n/2 -\log\frac{1}{\delta}$ as long as $m\geq n/2 + 2\log\frac{1}{\eps}.$
\end{theorem}

If $\eps$ is so low that the constraint $m\geq n/2+2\logeps$ is not satisfied,
then the construction can be modified as shown in Section~\ref{section-uniformity-dominates}.

\begin{proof}

\mypar{Extraction.} 
Our goal is to show that $R$ is nearly uniform given $P$. 
To do so, we first show that the function $h_i(a,b) = (\sigma, R)$
 is a universal hash family. Indeed, for $(a,b) \neq (a',b')$ consider 

\begin{eqnarray*}
\Pr_i[h_i(a,b) = h_i(a',b')]
&=&\Pr_i[ia+b = ia'+b']\\
&=&\Pr_i[i(a-a') = (b-b')]\\
&\leq&2^{-n/2}\,.   
\end{eqnarray*}
To see the last inequality recall that $(a,b) \neq (a',b').$
 Therefore, if $a = a'$, then $b\neq b'$ making
the $\Pr_i[i(a-a') = (b-b')]=0$. 
If $a \neq a'$, then there is a unique $i=(b-b')/(a-a')$ that
satisfies the equality. 
Since $i$ is chosen randomly from $\F_{2^{n/2}}$,
 the probability of the specific $i$ occurring 
 is $2^{-n/2}$.

Because $|(R, \sigma)|=n/2$,
Lemma~\ref{lemma-leftover-hash} gives us $\sd{(R,P),U_{|R|}\times U_{|P|}}\le \eps/2$
as long as $n/2\le m+2-2\log\frac{2}{\eps}$, or, equivalently, 
$(R,P)$ is $2^{(n/2-m)/2-1}$-close to $U_{|R|}\times U_{|P|}$. 
%We are interested in estimating the distance between
%the distributions $(R,P)$ and $U_{|R|}\times P$. 
Applying Lemma~\ref{lemma-sdswitch} to $A=R$, $B=P$, $C=U_{\frac{n}{2}-v}$,
$D=U_{\frac{n}{2}}\times U_{v}$, we get that $(R, P)$ is $\eps$-close
to $U_{(\frac{n}{2})-v}\times P$, for $\eps = 2^{(n/2-m)/2}$. 
From here it follows that for extraction to be possible, $m\geq n/2+ 2\logeps.$

\mypar{Post-Application Robustness.} 
In the post-application robustness security game, 
the adversary $\adv$ on receiving $(P=(i, \sigma), R)$ 
(generated 
according to procedure $\Gen$) outputs $P' = (i',\sigma')$, 
and is considered successful if $(P' \neq P) \wedge [i'a+ b]^v_1 = \sigma'$.
 In our analysis, we will assume that $i' \neq i$. We claim that this
 does not reduce $\adv$'s success probability.
Indeed, if $i' = i$ then, for $P'\neq P$ to hold, 
$\adv$ would have to output $\sigma' \neq \sigma$.
 However, when $i'=i$, $\Rep$ would output $\bot$ unless $\sigma' = \sigma$. 

In our analysis, we allow $\adv$ to be deterministic.
This is without loss of generality since we allow an unbounded adversary. 
We also allow $\adv$ to arbitrarily fix $i$. This 
makes the result only stronger since we demonstrate robustness for 
a worst-case choice of $i$. 

Since $i$ is fixed and $\adv$ is deterministic,
$(\sigma, R)$ determines 
the transcript  
$\tr=(i,\sigma,R,i',\sigma')$. 
For any particular $\tr$, let $\win_{\tr}$ be the event
that the transcript is $\tr$ and $\adv$ wins, i.e., that
$ia+b = \sigma||R \wedge [i'a+b]^v_1 = \sigma'$.
We denote by $\bad_{\tr}$
the set of $w=a||b$ that make $\win_{\tr}$
true.
For any $\tr$, $\Pr_w[\win_{\tr}] \leq |\bad_{\tr}|2^{-m}$, because each $w$
in $\bad_{\tr}$ occurs with probability at most $2^{-m}$.
We now partition the set $\bad_{\tr}$ into $2^{\ell}$ disjoint 
sets, indexed by $R' \in \zo^{\ell}$:
\begin{eqnarray*}
\bad^{R'}_{\tr} &\bydef& \{w\suchthat w\in \bad_{\tr} \wedge [i'a+b]^\ell_{v+1} = R'\}\\
&=& \{w\suchthat  (ia+b = \sigma||R) \wedge (i'a+b = \sigma'||R')\}
\end{eqnarray*}
\noindent
For a particular value of $(\tr,R')$, $w = a||b$ is uniquely 
determined by the constraints that define the above set 
\ifnum\lncs=0 Therefore,
\else i.e;
\fi 
$|\bad^{R'}_{\tr}| = 1$. Since 
$\bad_{\tr} = \bigcup_{R' \in \zo^\ell}\bad^{R'}_{\tr}$, we get
$|\bad_{\tr}| \le 2^{\ell} = 2^{{n/2}-v}$. 
 From here it follows that 
\[\Pr[\win_{\tr}] \le |\bad_{\tr}|2^{-m} \le 2^{{n/2}-v-m}\,.\] 

$\Pr[\win_{\tr}]$ measures the probability that
the transcript is $\tr$ and $\adv$ succeeds.
To find out the probability that $\adv$ succeeds,
we need to simply add $\Pr[\win_\tr]$ over all possible $\tr$.
Since a transcript is completely determined by $\sigma, R$,
the total number of possible transcripts is $2^{|\sigma|+|R|} = 2^{n/2}$
and, therefore,
$\adv$'s probability of success is at most $2^{n-v-m}$.

To achieve $\delta$-robustness, we need to set $v$ to at least $n-m+\logdel$. 
From here it follows that $\ell = \frac{n}{2} - v \leq \frac{1}{2}(2m - n  - 2\logdel)$. 
\ifnum\lncs=1\qed\fi
\end{proof}

\subsection{Getting Closer to Uniform}
\label{section-uniformity-dominates}
If $\eps$ is so low that the constraint $m\geq n/2+2\logeps$ is not satisfied,
then in our construction we can simply
shorten $R$ by $\beta = n/2+2\logeps-m$ bits,
as follows: 
keep  $v=n-m+\logdel$ (regardless of $\ell$), and let
$R=[ia+b]_{v+1}^{\ell+v}$, for any $\ell \le 2m-n-\logdel-2\logeps$.
This keeps $\sigma$ the same, but shortens $R$ enough for the leftover hash lemma to work.
The proof remains essentially the same,
except that to prove robustness, we will give the remaining bits $[ia+b]_{\ell+v+1}^{n/2}$
for free to $\adv$.

\subsection{Improving the construction of~\cite{DKRS06} When the Uniformity Constraint Dominates}
\label{section-dkrs-uniformity-dominates-improvement}
The construction of Dodis et al.~\cite{DKRS06} parses $w$ as two strings 
$a$ and $b$ of lengths $n-v$ and $v$, respectively. 
The values $\sigma,R$ are computed as $\sigma = [ia]^v_1+b$ and $R = [ia]^n_{v+1}$; 
$P = (i,\sigma)$. In order to get $R$ to be uniform given
$P$, the value $v$ is increased until the leftover hash lemma can be applied to $(R, \sigma)$.  However, we observe that this unnecessarily increases the length of $\sigma$ (i.e., for every bit added to $v$, two bits are subtracted from $R$).    Instead, we propose to improve this construction with essentially
the same technique as we use for our construction in Section~\ref{section-uniformity-dominates}.  The idea
is to simply shorten $R$  without
increasing the length of $\sigma$.
This improvement applies to both pre- and post-application robustness.  

For post-application robustness, suppose the uniformity constraint dominates, i.e.,
$2\logeps>(2m-n+\logdel)/3$.
Modify the construction of~\cite{DKRS06} by setting
$v=(2n-m+\logdel)/3$ and $R=[ia]_{v+1}^{n-v-\beta}$, where $\beta=2\logeps-(2m-n-\logdel)/3$.
This will result in an extracted key of length $\ell=(4m-2n-\logdel)/3-2\logeps$.  However,
even with the improvement, the extracted key will be always shorter than the key extracted by our scheme, as explained
in Section~\ref{section-dkrs-uniformity-dominates-comparison}

In contrast, this improvement seems useful in the case of pre-application robustness.
Again, suppose the uniformity constraint dominates, i.e.,
$2\logeps>\logdel$.
Modify the construction of~\cite{DKRS06} by setting $v = n - m +\logdel$ and 
$R = [ia]_{v+1}^{n-v-\beta}$, where $\beta=2\logeps-\logdel$. This will 
result in an extracted key of length $\ell=2m - n -2\logeps -\logdel$, which is
$2\logeps-\logdel$ longer than the key extracted without this modification.

\section{Comparison with the construction of \cite{DKRS06}}
\label{sec-comparison-dkrs}
\subsection{When the Robustness Constraint Dominates}
Recall that the construction of Dodis et al.~\cite{DKRS06} parses $w$ as two strings 
$a$ and $b$ of lengths $n-v$ and $v$, respectively. 
The values $\sigma,R$ are computed as $\sigma = [ia]^v_1+b$ and $R = [ia]^n_{v+1}$; 
$P = (i,\sigma)$. 
Notice that, like in our construction, increasing $v$ improves robustness and
decreases the number of extracted bits.  For pre-application robustness,
setting $v = n-m+\logdel$ suffices, and thus 
the construction extracts nearly $(2m-n)$ bits.
However, for post-application robustness, a much higher $v$ is needed, giving
only around $\frac{1}{3}(2m-n)$ extracted bits.

The post-application robustness game reveals more information to $\adv$ about
$w$ than the pre-application robustness game. This additional
information---namely, $R$---may make it easier for $\adv$ to guess
$\sigma'$ for a well-chosen $i'$.  
The key to our improvement is in the pairwise independence of the function
$ia+b$ that computes both $\sigma$ and $R$: because of pairwise independence,
the value $(\sigma, R)$ of the function 
on input $i$ tells $\adv$ nothing about the value $(\sigma',R')$ on another
input $i'$.  (This holds, of course, for uniformly chosen key $(a, b)$;
when $(a, b)$ has entropy $m$, then $\adv$ can find out $n-m$ bits of information
about $\sigma'$.)

In contrast, in the construction of~\cite{DKRS06}, only $\sigma$ is
computed using a pairwise independent hash function.  This works well
(in fact, better than our construction, because $b$ can be shorter)
for pre-application robustness, where $\adv$ does not find out $R$.
But it makes it possible for $R$ to decrease $\adv$'s uncertainty
about $\sigma'$ by as much as $\ell=|R|$, thus necessitating the
length $v$ of $\sigma'$ (and hence $\sigma$) to be $v>\ell+(n-m)$ (the
$(n-m)$ term is the amount of entropy already potentially ``missing''
from $\sigma'$ because of the nonuniformity of $w$).  See
Section~\ref{sec-dkrs-lowerbound} for a detailed description of an
adversarial strategy that utilizes $R$ to obtain $\sigma'$ in
the~\cite{DKRS06} construction.

Another way to see the differences between the two constructions is
through the proof.  In the proof of post-application robustness, the
transcript $\tr$ includes $R$, which makes for $2^\ell$ times more
transcripts than in the proof of pre-application robustness.  However,
the fact that this $R$ imposes an additional constraint of $w$, thus
reducing the size of the set $\bad_\tr$, can compensate for this
increase.  It turns out that for the construction of~\cite{DKRS06},
this additional constraint can be redundant if the adversary is clever
about choosing $i'$ and $\sigma'$, and the size of $\bad_\tr$ doesn't decrease.
Using a pairwise-independent function for computing $R$ in our
construction ensures that this additional constraint decreases the
size of $\bad_\tr$ by $2^\ell$.  Thus, our construction achieves the
same results for pre- and post-application robustness.

\subsection{When the Uniformity Constraint Dominates}
\label{section-dkrs-uniformity-dominates-comparison}
It should be noted that there may be reasonable cases when the
uniformity constraint $\eps$ on $R$ is strong enough that the
construction of \cite{DKRS06} extracts even fewer bits, because it
needs to take $v\ge n-m+2\logeps$ to ensure near-uniformity of $R$
given $P$. In that
case, as long as $m\ge n/2+2\logeps$, 
our construction will extract the same amount of bits as before,
thus giving it an even bigger advantage.
And when $m<n/2+2\logeps$,
our construction still extracts at least $3/2$ times more bits than the
construction of~\cite{DKRS06}, even with the improvement of
Section~\ref{section-dkrs-uniformity-dominates-improvement} applied
(this can be seen by algebraic manipulation of the relevant parameters
for the post-application robustness case).

\subsection{Why the construction of \cite{DKRS06} cannot extract more bits}
\label{sec-dkrs-lowerbound}
Recall that the robust fuzzy extractor of~\cite{DKRS06} operates as
follows: parse $w$ as two strings $a, b$ of lengths $n-v,v$
respectively and compute $\sigma =
[ia]^v_1+b$ and $R = [ia]^n_{v+1}$; $P = (i,\sigma)$.

For post-application robustness, the concern is that $R$ can reveal
information to the adversary about $\sigma'$ for a cleverly chosen
$i'$.  Because the length of $\sigma'$ is $v$ and $\ell+(n-m)$ bits of
information about $\sigma'$ may be available (the $\ell$ term comes
from $|R|$, and $(n-m)$ term comes from the part of $w$ which has no
entropy), this leads to the requirement that $v\ge \ell+n-m+\logdel$
to make sure the adversary has to guess at least $\logdel$ bits about
$\sigma'$.  Plugging in $\ell=n-2v$, we obtain $\ell\le
\frac{2}{3}(m-n/2-\logdel)$, which is the amount extracted by the
construction.

Here we show an adversarial strategy that indeed utilizes $R$ to
obtain information about $\sigma'$ to succeed with probability
$\delta/2$.  This demonstrates that the analysis in~\cite{DKRS06} is
tight up to one bit.  To do so we have to fix a particular (and
somewhat unusual) representation of field elements.  (Recall that any
representation of field elements works for constructions here and
in~\cite{DKRS06}, as long as addition of field elements corresponds to
the exclusive-or of bit strings.)  Typically, one views $\F_{2^{n-v}}$
as $\F_2[x]/(p(x))$ for some irreducible polynomial $p$ of degree
$n-v$, and represents elements as $\F_2$-valued vectors in the basis
$(x^{n-v-1}, x^{n-v-2}, ..., x^2, x, 1)$.  We will do the same, but
will reorder the basis elements so as to separate the even and the odd
powers of $x$: 
\ifnum\lncs =0 $(x^{n-v-1}, x^{n-v-3}, \dots, x, x^{n-v-2}, x^{n-v-4},
\dots, 1)$
\else $(x^{n-v-1}, x^{n-v-3}, \dots, x, \allowbreak x^{n-v-2}, x^{n-v-4},
\dots, 1)$
\fi
(assuming, for concreteness, that $n-v$ is even).  The
advantage of this representation for us is that the top half of
bits of some value $z\in F_{2^{n-v}}$ is equal to the bottom half of the
bits of $z/x$, as long as the last bit of $z$ is 0.

Now suppose the distribution on $w$ is such that the top $n-m$ bits of
$b$ are 0 (the rest of the bits of $w$ are uniform).  Then by
receiving $\sigma$ and $R$, the adversary gets to see the top
$\ell+(n-m)$ bits of $ia$.  Therefore, the adversary knows
$\ell+(n-m)$ bits from the bottom half of $ia/x$ as long as the last
bit of $ia$ is 0, which happens with probability $1/2$.  To use this
knowledge, the adversary will simply ensure that the difference
between $\sigma'$ and $\sigma$ is $[ia/x]_1^v$, by letting $i'=i+i/x$.

Thus, the adversarial strategy is as follows: let $i'=i+i/x$; let
$\tau$ consist of the $\ell$ bits of $R$, the top $n-m$ bits of
$\sigma$, and $\logdel=v-\ell-(n-m)$ randomly guessed bits, and let
$\sigma'=\sigma+\tau$.  The adversary wins whenever $\tau=[ia/x]_1^v$,
which happens with probability $2^{v-\ell-(n-m)}/2=\delta/2$, because all but
$\logdel$ bits of $\tau$ are definitely correct as long as the last
bit of $ia$ is 0.

The above discussion  gives us the following result.
\begin{theorem}
\label{theorem-dkrs-lowerbound}
There exists a basis for $GF(2^{n-v})$ such that
for any integer $m$ there exists a distribution $W$ of min-entropy $m$ for which
the post-application robustness of the construction from \cite[Theorem 3]{DKRS06}
can be violated with probability at least $\delta/2$,
where $v$ is set as required for robustness $\delta$
by the construction (i.e., $v=(n-\ell)/2$ for $\ell=(2m-n-2\logdel)/3$).
\end{theorem}

Note that our lower bound uses a specific representation of field
elements, and hence does not rule out that for some particular
representation of field elements, a lower value of $v$ and, therefore,
a higher value of $\ell$ is possible.  However, a security proof for a
lower value of $v$ would have to then depend on the properties of that
particular representation and would not cover the
construction of~\cite{DKRS06} in general.

\section{Tolerating Binary Hamming Errors}
\label{sec-rfe-errors} 

We now consider the scenario where Bob has 
a string $w'$ that is close to Alice's input $w$ (in the Hamming metric). 
In order for them to agree on a random string, Bob 
would first have recover $w$ from $w'$. To 
this end, Alice could send the secure sketch $s=\ff(w)$ 
to Bob along with $(i,\sigma)$. 
To prevent an undetected modification of $s$ to $s'$, she could 
send an authentication of $s$ (using $w$ as the key) as well. 
The nontriviality of making such an extension work  
arises from the fact that modifying $s$ to $s'$ also
gives the adversary the power to influence
Bob's verification key $w^* = \ssrec(w',s')$. The adversary 
could perhaps exploit this circularity 
to succeed in an active attack (the definition of 
standard authentication schemes only guarantee security 
when the keys used for authentication and verification are the same). 

We break this circularity by exploiting 
the algebraic properties of the Hamming metric space,
and using authentication secure against algebraic
manipulation~\cite{DKRS06,CDFPW08}. 
The techniques that we use are essentially the same as
used in ~\cite{DKRS06}, but adapted to our construction.
We present the construction here and then 
discuss the exact properties that we use in the proof of security.

\mypar{Construction.}
Let $\M$ be the Hamming metric space on $\zo^n$. 
Let $W$ be a distribution of min-entropy $m$ over $\M$. 
Let $s = \ff(w)$ be a deterministic, 
linear secure sketch; let $|s| = k$, $n' = n-k$. Assume that $\ff$ is a surjective linear 
function (which is the case for the syndrome construction for the Hamming metric mentioned in Section~\ref{section-prelims}). Therefore, 
there exists a $k \times n$ matrix $\sS$ of rank $k$ such that $\ff(w) = \sS w$. 
Let $\sS^\bot$ be an $n' \times n$ matrix such that
 $n \times n$ matrix $\left(\frac{\sS}{\sS^\bot}\right)$ 
has full rank. We let $\ff^\bot(w) = \sS^\bot(w)$.  

To compute $\Gen(w)$, let $s = \ff(w)$, $c = \ff^\bot(w)$; $|c| = n'$.  
We assume that $n'$ is even (if not, drop one bit of $c$,  reducing its entropy 
by at most 1). Let $a$ be the first half of $c$ and $b$ the second. View $a,b$ 
as elements of $\F_{2^{n'/2}}$. Let $L = 2\lceil \frac{k}{n'} \rceil$ (it will
important for security that $L$ is even).
Pad $s$ with 0s to length $Ln'/2$, and then
split it into $L$ bit strings $s_{L-1}, \dots, s_0$
of length $n'/2$ bits each, viewing
each bit string as an element of $\F_{2^{n'/2}}$.
 Select $i\leftarrow \F_{2^{n'/2}}.$
Define $f_{s,i}(x) = x^{L+3}+x^2(s_{L-1}x^{L-1} + s_{L-2}x^{L-2}+\dots+s_0) + ix.$ 
Set $\sigma = [f_{s,i}(a)+b]^{v}_{1}$, and output
$P=(s,i,\sigma)$ and $R= [f_{s,i}(a)+b]^{n'/2}_{v+1}$.

\begin{tabbing}
\ifnum\lncs=0
\\
wwwwwwwwwwww\=www\=3. \=wwwwwwwww\=\kill
\else
ww\=wwww\=3. \=wwwwwwwww\=\kill
\fi
\>\+
$\Gen(w)$:\\
1. Set $s = \ff(w)$, $c = \ff^\bot(w)$, $k = |s|$, $n' =|c|$. \\
\> - Let $a = [c]^{n'/2}_1$, $b=[c]^{n'}_{n'/2+1}$\\
\> - Let $L = 2\lceil \frac{k}{n'} \rceil.$ Pad $s$ with 0s to length $Ln'/2$.\\
\> - Parse the padded $s$ as $s_{L-1}||s_{L-2}|| \ldots || s_0$ for $s_i\in \F_{2^{n'/2}}$.\\
2. Select $i\leftarrow \F_{2^{n'/2}}.$\\
%\> - Define $f_{s,i}(x) = x^{L+3}+x^2(s_{L-1}x^{L-1} + s_{L-2}x^{L-2}+\dots+s_0) + ix$\\
3. Set $\sigma = [f_{s,i}(a)+b]^{v}_{1}$, and output $R= [f_{s,i}(a)+b]^{n'/2}_{v+1}$ and $P=(s,i, \sigma).$\\ 
\ \\
$\Rep(w',P' = (s',i',\sigma'))$:\\
1. Compute $w^* = \ssrec(w',s')$\\
\> - Verify that $\disfn(w^*,w') \leq t$ and $\ff(w^*) = s'$. If not, output $\bot$. \\
2. Let $c'=\ff^{\perp}(w^*)$. Parse $c'$ as $a'||b'$. \\
3. Compute $\sigma^* = [f_{s',i'}(a')+b']^{v}_{1}$. \\
\> - Verify that $\sigma^* = \sigma'$. If so, output $R= [f_{s',i'}(a')+b']^{n'/2}_{v+1}$, else output $\bot$. 
\end{tabbing}

In the theorem statement below, let $B$ denote the volume of a Hamming ball or radius $t$ in $\zo^n$ ($\log B\le nH_2(t/n)$~\cite[Chapter 10, \S 11, Lemma 8]{MW77} and $\log B \le t \log (n+1)$~\cite{DKRS06}).

\begin{theorem}
Assume $\ff$ is a deterministic linear $(m,m-k,t)-$secure sketch of output length $k$ for the Hamming metric 
on $\zo^n$. Setting $v =(n-k)/2-l$, the above construction 
is an $(m,l,t,\eps)$ fuzzy extractor with robustness $\delta$ for any $m,l,t,\eps$ satisfying
$l\leq m-n/2-k-\log B - \log\left(2\left\lceil\frac{k}{n-k}\right\rceil +2\right)-\logdel$
as long as $m\geq \frac{1}{2}(n+k)+2\logeps.$ 
\end{theorem}

Again, if $m<\frac{1}{2}(n+k)+2\logeps$, the construction can be modified,
as shown in Section~\ref{section-closer-to-uniform-fuzzy}.

\begin{proof}
\mypar{Extraction.} 
Our goal is to show that $R$ is nearly uniform given $P = (i,s,\sigma)$. 
To do so, we first note that for every $s$, 
the function $h_i(c) = (\sigma, R)$ 
is a universal hash family. Indeed for $c \neq c'$ there is a unique $i$ such that $h_i(c) = h_i(c')$ 
(since $i(a-a')$ is fixed, like in the errorless case). 
We also note that $\thinf(c\mid \ff(W)) \ge \thinf(c, \ff(W))-k=\hinf(W)-k=m-k$
by Lemma~\ref{lemma-avg-min-entropy}. Because $|(R,\sigma)| = n'/2$,
Lemma~\ref{lemma-leftover-hash} (or, more precisely,
its generalization mentioned in the paragraph following the lemma,
needed here because $h_i$ depends on $s$)
gives us  
\[\sd{(R,P),U_{|R|} \times \ff(W) \times U_{n'/2}\times U_{v}}\le \eps/2\] for 
$n'/2 \leq m -k +2 -2\log(2/\eps)$.  This is equivalent to saying that 
$(R,P)$ is $2^{(n'/2-m+k)\frac{1}{2}-1}$-close to $U_{|R|} \times \ff(W) \times U_{n'/2}\times U_{v}$. 

Applying Lemma~\ref{lemma-sdswitch} to $A=R$, $B=P$, $C=U_{n'/2-v}$,
$D=\ff(w)\times U_{n'/2}\times U_v$, we get that $(R, P)$ is
$\eps$-close
to $U_{\frac{n'}{2}-v}\times P$, for $\eps =
2^{(\frac{n'}{2}-m+k)/2}.$ 

From here it follows that for extraction to be possible,  $m\geq \frac{1}{2}(n+k)+2\logeps\,.$   

\mypar{Post-Application Robustness.} 
In the post-application robustness security game, 
the adversary $\adv$ on receiving $(P=(s,i, \sigma), R)$ 
(generated according to procedure $\Gen$) outputs
$P' = (s',i',\sigma')$,  
and is considered successful if 
$(P' \neq P) \wedge \Rep(w',s') \neq \bot$. 
In our analysis, we will assume that $(i',s') \neq (i,s)$. 
We claim that this does not reduce $\adv$'s success probability.
Indeed, if $(i',s') = (i,s)$ then, 
$c'$ computed within $\Rep$ will equal $c$. 
So, for $P'\neq P$ to hold, 
$\adv$ would have to output $\sigma' \neq \sigma$. 
However, when $(i',c' ,s')=(i,c,s)$, $\Rep$ would compute
$\sigma^*=\sigma$, and therefore
would output $\bot$ unless $\sigma' = \sigma$.

In our analysis, we allow $\adv$ to be deterministic.
This is without loss of generality since we allow an unbounded adversary. 
We also allow $\adv$ to arbitrarily fix $i$. This  
makes the result only stronger since we demonstrate robustness for 
a worst-case choice of $i$. 

Since $i$ is fixed and $\adv$ is deterministic,
the $\tr = (i,s,\sigma,R,i',s',\sigma')$ 
is determined completely by $(s,\sigma, R)$. 
Recall that the prime challenge in constructing 
a robust fuzzy extractor was that $\adv$ 
could somehow relate the key used by $\Rep$ to verify $\sigma'$ to the
authentication key that was used by $\Gen$ to come up with $\sigma$.
As was done in ~\cite{DKRS06}, 
we will argue security of our construction 
by showing that the 
MAC
scheme implicitly used in our construction remains unforgeable 
even when $\adv$ could force the verification 
key to be at an offset (of her choice) from the authentication key. 
We will formalize such an argument by assuming that $\adv$
learns $\Delta = w' - w$. Recall that $w^* = \ssrec(w',s')$ and $c'=a'||b'=\ff^{\perp}(w^*) $. 
The following claim that was proven in~\cite{DKRS06} 
states that given $(\Delta, s)$, $\adv$ can compute the 
offsets $\Delta_a =a'-a, \Delta_b =b'-b$ induced 
by her choice of $s'$. 

\begin{claim} 
Given $\Delta = w'-w,$  and the sketches $s,s',\adv$ can compute $\Delta_a = a' - a$ and 
$\Delta_b = b' - b$, or determine that $\Rep$ will reject before computing $a',b'$.
\end{claim}

In other words, she can compute the 
offset between the authentication key that 
$\Gen$ used to come up with $\sigma$ 
and the verification key that $\Rep$ will use 
to verify $\sigma'$. We will now argue that as long as $W$ has sufficient
min-entropy, even knowing the offset does not help $\adv$ succeed 
in an active attack. Recall that since $i$ is arbitrarily fixed by $\adv$, 
$\adv$'s success depends on $w,w'$, or, alternatively, on $w,\Delta$. 
Fix some $\Delta$.  For any particular $\tr$, 
let $\win_{\tr,\Delta}$ be the event that 
the transcript is $\tr$ and $\adv$ wins, 
i.e., that 
$f_{s,i}(a)+b = \sigma||R \wedge [f_{s',i'}(a')+b']^v_1 = \sigma' \wedge \ff(w)=s$,
conditioned on the fact that $w'-w$ is $\Delta$.
We denote by $\bad_{\tr,\Delta}$
the set of $w$ that make $\win_{\tr,\Delta}$
true. We now partition the set $\bad_{\tr,\Delta}$ into $2^{\ell}$ disjoint 
sets, indexed by $R' \in \zo^{\ell}$:
\begin{eqnarray*}
\bad^{R'}_{\tr,\Delta} &\bydef& \{w\suchthat w\in \bad_{\tr,\Delta} \wedge [f_{s',i'}(a')+b']^\ell_{v+1} = R'\}\\
&=& \{w\suchthat  (f_{s,i}(a)+b = \sigma||R) \wedge (f_{s',i'}(a')+b' = \sigma'||R')\wedge \ff(w)=s\}.
\end{eqnarray*}
By Claim 1, fixing $(\tr,\Delta)$, also fixes $\Delta_a,\Delta_b$. 
It follows that every $w\in \bad^{R'}_{\tr,\Delta}$ needs to satisfy  
\[f_{s,i}(a) - f_{s',i'}(a+\Delta_a) = (\Delta_b+\sigma-\sigma')||(R-R') \wedge \ff(w)=s.\]
For a given $\tr,\Delta,R'$, the right hand side of the first equation takes a fixed value. 
Let us now focus on the polynomial $f_{s,i}(a) - f_{s',i'}(a+\Delta_a)$. 
We will consider two cases:
\begin{itemize}
\item $\Delta_a = 0$: In this case, 
$f_{s,i}(x) - f_{s',i'}(x)$ is a polynomial in which 
a coefficient of degree 2 or higher is nonzero if $s \neq s'$ and 
a coefficient of degree 1 or higher is nonzero if $i \neq i'$. 
\item $\Delta_a \neq 0$: 
Observe that the leading term of the polynomial is $((L+3)\bmod 2)\Delta_ax^{L+2}$. 
Since we forced $L$ to be even, the coefficient of the leading term is 
nonzero, making $f_{s,i}(x) - f_{s',i'}(x+\Delta_a)$
a polynomial of degree $L + 2$.  
\end{itemize}
Therefore, in either case, the $f_{s,i}(x) - f_{s',i'}(x+\Delta_a)$ 
is a nonconstant polynomial of degree at most $L+2$. 
A nonconstant polynomial of degree $d$ can take on a fixed value
at most $d$ times. It, therefore, follows that there are at most 
$L+2$ values of $a$ such that 
$f_{s,i}(a) - f_{s',i'}(a+\Delta_a) = (\Delta_b+\sigma-\sigma')||(R-R')$. 
Each such $a$ uniquely determines $b=(\sigma||R)-f_{s,i}(a)$. And $w$ is uniquely determined by $c=a||b=\ff^{\bot}(w)$ and $s=\ff(w)$.
Therefore, there are at most $L+2$
values of $w$ in the set $\bad^{R'}_{\tr,\Delta}$ i.e, $|\bad^{R'}_{\tr,\Delta}| \leq L+2$. 
Since $\bad_{\tr,\Delta} = \bigcup_{R' \in \zo^\ell}\bad^{R'}_{\tr,\Delta}$, we get 
$|\bad_{\tr,\Delta}| \le (L+2)2^{\ell} = (L+2)2^{{n'/2}-v}$. 
Thus, $\Pr_w[\win_{\tr,\Delta}] \leq |\bad_{\tr}|2^{-\hinf(w|\Delta)} \leq (L+2)2^{n'/2-v-\hinf(w|\Delta)}$.  

To find out the probability $\Pr_w[\win_\Delta]$
that $\adv$ succeeds conditioned on a
particular $\Delta$, we need to add up $\Pr_w[\win_{\tr,\Delta}]$ over
all possible transcripts.  Recalling that each transcript is determined
by $\sigma, R$ and $s$ and hence there are $2^{n'/2+k}$ of them, and that
$n'+k=n$,
we get $\Pr_w[\win_\Delta]\leq (L+2)2^{n-v-\hinf(w|\Delta)}$.  

Finally, the probability of adversarial success it at most
\[
\expe_{\Delta}\Pr_w[\win_\Delta] \le  (L+2)2^{n-v-\thinf(w|\Delta)}\,.
\]
In particular, if the errors $\Delta$ are independent of $w$,
then $\thinf(w|\Delta)=\hinf(w)=m$, and the probability of adversarial success is at most $(L+2)2^{n-v-m}$.  
In the worst case, however, the entropy of $w$ may decrease at most by the number of bits needed to represent $\Delta.$
Let $B$ be the volume of the hamming ball of radius $t$ in $\zo^n$. Then, $\Delta$ can be represented in 
$\log B$ bits and $\thinf(w|\Delta) \geq m - \log B$, by Lemma~\ref{lemma-avg-min-entropy}. From here it follows that 
\[
\Pr[\A's\ success] \leq B(L+2)2^{n-v-m}
\]
To achieve $\delta-$robustness, we want  $B(L+2)2^{n-v-m} \leq \delta$ 
i.e.,  $v \geq n-m +\log B +\log(L+2) + \logdel$. Setting $v = n-m+\log B+\log(L+2) +\logdel$, 
and using $L = 2\lceil\frac{k}{n-k}\rceil$
it follows that \[\ell \leq m - n/2 - k -\log B - \log\left(2\left\lceil\frac{k}{n-k}\right\rceil+2\right)  - \logdel\,. \]
\ifnum\lncs=1\qed\fi
\end{proof}

\subsection{Getting Closer to Uniform}
\label{section-closer-to-uniform-fuzzy}

If $\eps$ is so low that $m \ge \frac{1}{2}(n+k) + 2\logeps$ does not hold,
we can modify our construction 
just as we did in section~\ref{section-uniformity-dominates}, by
shortening $R$ by $\beta=\frac{1}{2}(n+k)+2\logeps-m$.
That is,  keep  $v = n-m +\log B+\log(L+2)+\logdel$ fixed and 
let $R = [f_{s,i}(a)+b]_{v+1}^{\ell+v}$, where
$\ell \le 
n/2 -v- \beta$. 

\ifnum\lncs=0
\section*{Acknowledgements}
This work was supported in part by the U.S. National Science Foundation grants CCF-0515100 and CNS-0546614.
\fi

%%%%%%%%%%%%%%%%%%%%%%%%%%%%%%%%%%%%%%%%%%%%%%%%%%%%%%
%\bibliographystyle{alpha}
%\bibliography{bib/fuzzy}

\newcommand{\etalchar}[1]{$^{#1}$}

\appendix

\iffalse

\fi

\end{document}